\documentclass[prd,aps,superscriptaddress,floatfix,nofootinbib,twocolumn]{revtex4-1}
\pdfoutput=1
\usepackage{amsmath}
\usepackage{amssymb}
\usepackage{graphicx}
\usepackage{color}

\def\Fig#1{Fig.~\ref{#1}} 

\def\eq#1{(\ref{#1})}

\begin{document}

\title{The phase structure of QCD for heavy quarks}

\author{Christian S. Fischer} 
\affiliation{Institut f\"ur Theoretische
  Physik, Justus-Liebig-Universit\"at Gie\ss{}en, Heinrich-Buff-Ring
  16, D-35392 Gie\ss{}en, Germany.}  
\author{Jan Luecker}
\affiliation{Institut f\"ur Theoretische Physik, Goethe-Universit\"at
  Frankfurt, Max-von-Laue-stra\ss{}e 1, D-60438 Frankfurt/Main,
  Germany} 
\affiliation{Institut f\"{u}r Theoretische Physik,
  Universit\"{a}t Heidelberg, Philosophenweg 16, 69120 Heidelberg,
  Germany} 
\author{Jan M.~Pawlowski} \affiliation{Institut f\"{u}r
  Theoretische Physik, Universit\"{a}t Heidelberg, Philosophenweg 16,
  69120 Heidelberg, Germany}

\date{\today}
\begin{abstract}
  We investigate the nature of the deconfinement and Roberge-Weiss
  transition in the heavy quark regime for finite real and imaginary
  chemical potential within the functional approach to continuum
  QCD. We extract the critical phase boundary between the first order
  and cross-over regions, and also explore tricritical scaling. Our
  results confirm previous ones from finite volume lattice studies.
\end{abstract}

\maketitle

\section{Introduction}

The phase structure of QCD at finite temperature and density is a very
active research topic explored experimentally in heavy ion collisions
at RHIC, the LHC and the future NICA and FAIR facilities. One of the
most interesting problems concerns the possible appearance of a
critical end point at finite chemical potential $\mu$, connecting the
chiral and deconfinement crossover region at small $\mu$ with the
first order transition at large chemical potential. In this region of
the phase diagram, lattice QCD is severely hampered by the fermion
sign problem preventing continuum extrapolated studies with
contemporary resources, see e.g. \cite{Philipsen:2010gj,Aarts:2013bla}
for reviews. First principle continuum methods such as
the approach via Dyson-Schwinger and functional renormalization group
equations avoid this problem at the expense of truncations, that need
to be controlled. At finite real and imaginary chemical potential such
control is possible in the heavy quark region, which provides an
interesting playground with interesting physical phenomena well worth
studying in their own right.

\begin{figure}[b]
\includegraphics[width=.33\textwidth]{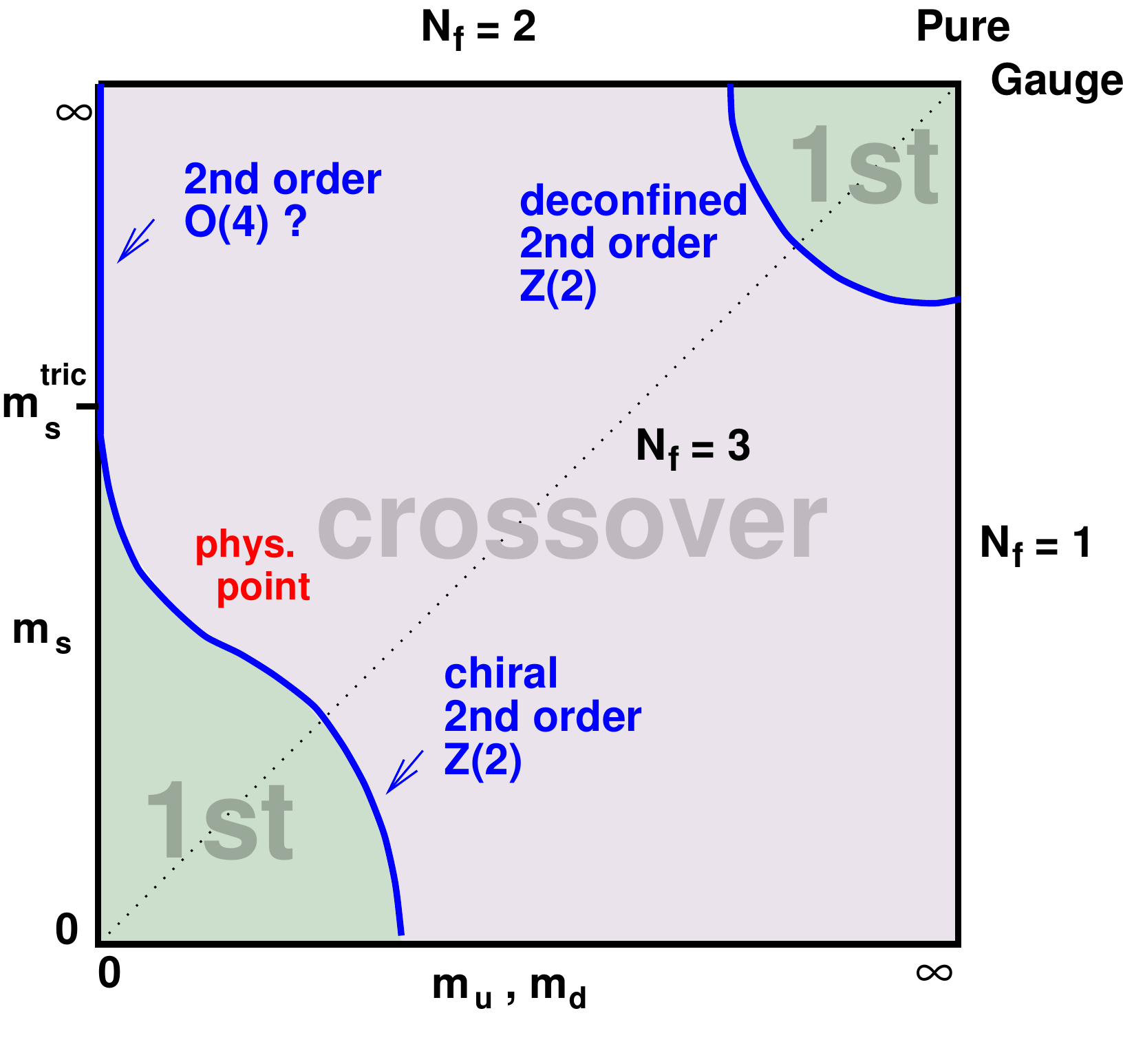}
\caption{The Columbia plot, taken from \cite{deForcrand:2010he}. \label{fig:ColumbiaPlot}}
\end{figure}

At zero chemical potential and infinite quark mass, i.e. in the
quenched approximation of QCD, the Polyakov loop expectation value
exhibits a discontinuity which is associated with a first order
deconfinement transition.  However, due to the center-symmetry
breaking effect of dynamical quarks, this transition is known to be a
crossover for physical quark masses 
\cite{Aoki:2006we,Bazavov:2011nk,Bhattacharya:2014ara}. 
It follows that a critical quark
mass exists, where the first order phase transition changes into a
crossover.  This situation is usually represented in the upper right
corner of the Columbia plot, see Fig.~\ref{fig:ColumbiaPlot}.
This regime of QCD has been studied in lattice simulations, where due
to the large quark masses the sign problem becomes treatable, and
hence the extension to finite chemical potential is also
feasible \cite{Alexandrou:1998wv,Saito:2011fs,Fromm:2011qi,Saito:2013vja}. 
Then the critical line extents to a critical surface, which
has been studied for real and imaginary chemical potential
\cite{Fromm:2011qi}. QCD at (unphysical) imaginary chemical potential
does not suffer from a sign problem even a small quark masses, as
$\gamma_5$ hermiticity is restored.  Such an extension is of high
interest as it allows to study QCD in a fugacity expansion. Moreover,
the study of the analytic structure of the critical surface at large
quark masses may give hints at how to extend the critical surface at
small quark masses from imaginary chemical potential to real chemical
potential.

In this work we apply the continuum Dyson-Schwinger approach to QCD in
the heavy quark limit. Our approximation scheme explicitly takes into
account the back-reaction of the quarks onto the Yang-Mills sector
thus rendering a systematic exploration of the physics of the Columbia
plot possible \cite{Fischer:2012vc,Fischer:2014ata}. We are therefore
able to calculate the Polyakov-loop potential explicitly using a DSE
for the background gauge field \cite{Fister:2013bh,Fischer:2013eca}.
For applications within functional approaches see
\cite{Braun:2009gm,Stiele:2013gra,Herbst:2014zea,Fukushima:2012qa,Reinosa:2014ooa}.
This approach allows us to study the physics of the deconfinement
transition directly without the need for any {\it ans\"{a}tze} for the
potential as has been used in model studies, e.g.\
\cite{Kashiwa:2012wa,Lo:2014vba}.  In this framework we explore the
heavy quark regime and determine the critical surface of the
deconfinement transition. In our continuum approach we find
tricritical scaling in agreement with the finite volume lattice
studies.

This letter is organized as follows: In the next section we present
some technical details related to our approach. We then discuss our
results for zero, real and imaginary chemical potential in section
\ref{results} before we conclude in section \ref{conclusions}.

\section{Set-up}

\subsection{Polyakov-loop potential}

\begin{figure}
\includegraphics[width=.45\textwidth]{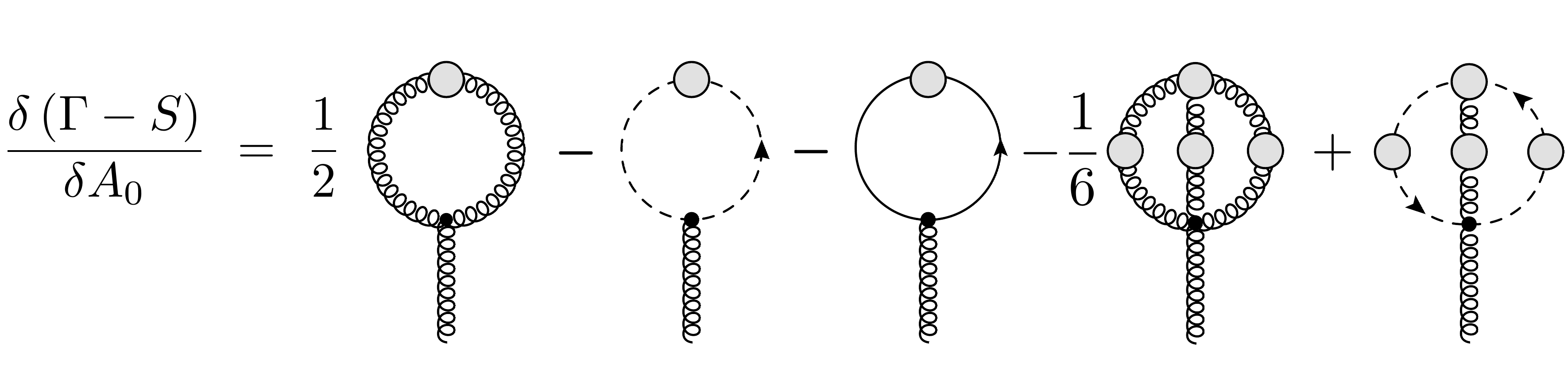}
\caption{The DSE for background field $\bar{A}$. \label{fig:DSE-A}}
\end{figure}

In the heavy quark limit of QCD, confinement is associated to center
symmetry.  The expectation value of the Polyakov loop, $\langle L[A_0]
\rangle$, serves as an order parameter for center symmetry. It is
linked to the free energy of a quark--anti-quark pair at infinite
distance. $\langle L[A_0] \rangle$ is the
minimum of the order parameter potential $\Omega(L)$. This potential is often
used as input in enhanced model studies. An alternative order parameter is $L[\bar
A_0 ]$, with background field $\bar A_0$, that satisfies the quantum equation
of motion \cite{Braun:2007bx,Marhauser:2008fz}. The two are related by
\begin{equation}
L[\bar{A_0}] \ge \langle L[A_0] \rangle\,
\quad {\rm and}\quad \langle L[A_0] \rangle=0\to 
L[\bar{A_0}] =0\,,
\label{eq:jensen+0}\end{equation}
where
\begin{equation}
  L[A_0] = \frac{1}{N_c}\mathrm{Tr}_c {\cal P} 
  \left[e^{i\,g\,\int_0^{\beta}dx_0\,A_0(x_0,\vec{x})}\right],
\end{equation}
is the Polyakov loop for a gauge field $A$, ${\cal P}$ stands for path
ordering and $\beta=1/T$ is the inverse temperature. The two relations
in \eq{eq:jensen+0} mark $L[\bar{A_0}]$ as an order parameter.

In the theory with background field, the DSE for this field has been
derived in \cite{Fister:2013bh}, see Fig.~\ref{fig:DSE-A}. This DSE
describes the derivative of the background-field potential
$V(\bar{A}_0)=\Omega(
L)$. From the minimum of the potential, we
get the order parameters $\bar{A}_0$ and $L[\bar A_0]$.

Constant background fields can always be rotated in the Cartan sub-algebra of the
$SU(3)$ color group. In the fundamental representation the background field is
then decomposed to
\begin{equation}
\bar{A}_0 = \frac{2\pi T}{g}\left(\varphi_3 \frac{\lambda_3}{2} + \varphi_8 \frac{\lambda_8}{2}\right),
\label{eq:bfDecomposition}
\end{equation}
with Gell-Mann matrices $\lambda_a$. The Polyakov loop $L[\bar A_0]$ reads 
\begin{equation}
L[\bar A_0]
  = \frac{1}{3}\left[e^{-i\frac{2\pi\varphi_8}{\sqrt{3}}}
	+ 2e^{-i\frac{\pi\varphi_8}{\sqrt{3}}}\cos(\pi\varphi_3)\right],
\end{equation}
a complex function, which is real for $\varphi_8=0$. 

In \cite{Fischer:2013eca} the background-field DSE has been used to
study the Polyakov loop potential for QCD at finite density, and we
will build upon this work.  It has been argued in \cite{Fister:2013bh}
that the two-loop terms in Fig.~\ref{fig:DSE-A} can be neglected in an
optimized regularization scheme and for temperatures not too far away
from the critical temperature. In the present work we are interested
in the physics close to the critical surface and hence the two-loop
terms can be safely dropped. Then the Polyakov-loop potential is
completely determined by the quark, gluon and ghost propagators.

\subsection{Propagators}
\begin{figure}
\includegraphics[width=.45\textwidth]{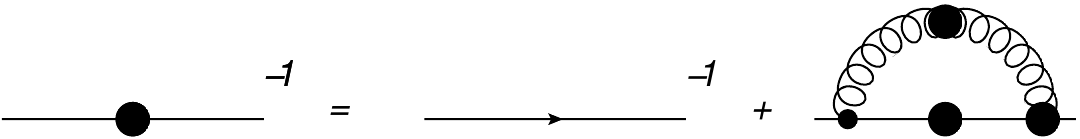}
\includegraphics[width=.42\textwidth]{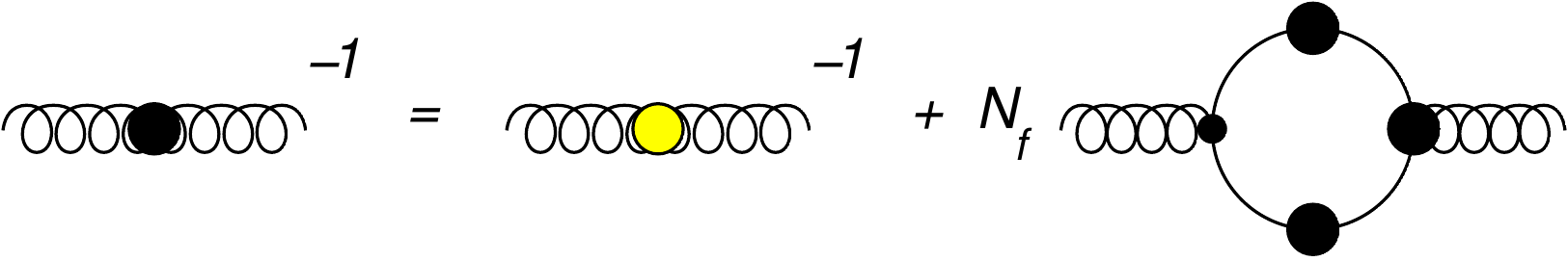}
\caption{The DSEs for quark and gluon propagators. \label{fig:qgDSEs}}
\end{figure}
It is left to determine gluon, ghost and quark propagators. The
computation is based on the quenched ghost and gluon propagators at
finite temperature obtained within the functional renormalisation
group approach in \cite{Fister:2011uw}. We use the quenched ghost in
the following: at vanishing density both thermal and quantum
fluctuations of the matter sector have a negligible impact on the
ghost propagator, it keeps its quenched form. As density fluctuations
are transmitted via the matter sector, this behaviour can be extended
to finite density. In contrast, it is mandatory for qunatitative
precision to unquench the gluon.  To this end, we solve quark and
gluon DSEs in a truncation that has been developed in
\cite{Fischer:2012vc,Fischer:2014ata}; the truncated DSEs are shown in
Fig.~\ref{fig:qgDSEs}.

The bare quark propagator is given by
\begin{equation}
S_0^{-1}(p) = i(\omega_n+i\mu)\gamma_4+i\vec{p}\vec{\gamma}+Z_m m,
\end{equation}
where $m(\zeta)$ is the renormalized bare quark mass at renormalization point $\zeta$ 
and $Z_m(\zeta)$ its renormalization factor. For the calculation we use $\zeta = 80$ GeV.
The quark mass is the main parameter, which we will tune in order 
to find its critical value $m_c$, where the phase transition is of second order.

For the quark-gluon vertex we choose the same ansatz that has been shown to
give excellent results in comparison with lattice QCD at small quark masses 
in Ref.~\cite{Fischer:2012vc,Fischer:2014ata}. The only change is related to
the infrared strength of the quark-gluon interaction controlled by a 
parameter $d_1$. In \cite{Williams:2014iea}, the DSE for the quark-gluon 
vertex has been solved at zero temperature in a truncation that allowed to 
extract the quark mass dependence of the vertex. Guided by these results,
we infer a reduction of the vertex strength from $d_1 \approx 8$ GeV$^2$ to
the value $d_1=0.5$ GeV$^2$ used in this study.

\section{Determination of critical quark masses}\label{results}

\subsection{Zero density}

\begin{figure}
\includegraphics[width=.45\textwidth]{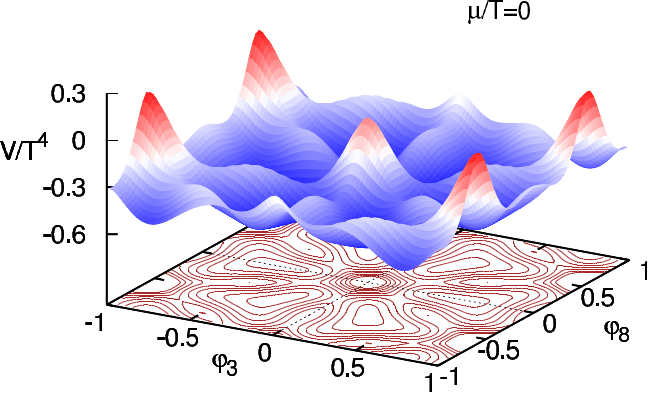}
\caption{Polyakov-loop potential at $\mu=0$ in the approximately center symmetric
  phase. \label{fig:pot2dmu0}}
\end{figure}

We show the Polyakov-loop potential at $\mu=0$ in Fig.~\ref{fig:pot2dmu0} as a
function of $\varphi_3$ and $\varphi_8$ for $T=250$ MeV and $m=320$ MeV in the
approximately center-symmetric phase. Here, the potential has six degenerate
minima located close to $(\varphi_3,\varphi_8) = (\pm 2/3,0)$ and
$(\varphi_3,\varphi_8) = (\pm 1/3, \pm 1/\sqrt{3})$ rendering $L[\bar A_0]\approx0$.
Consider now the minimum at $(\varphi_3,\varphi_8) = (2/3,0)$. At the critical
temperature $T_c$ a first order phase transition can clearly be distinguished
from a crossover by the emergence of a two-minima structure of the potential as
a function of $\varphi_3$ around $T_c$. At the critical quark mass $m_c$ the
two-minima structure changes to a single minimum.
\begin{figure}
\includegraphics[width=.45\textwidth]{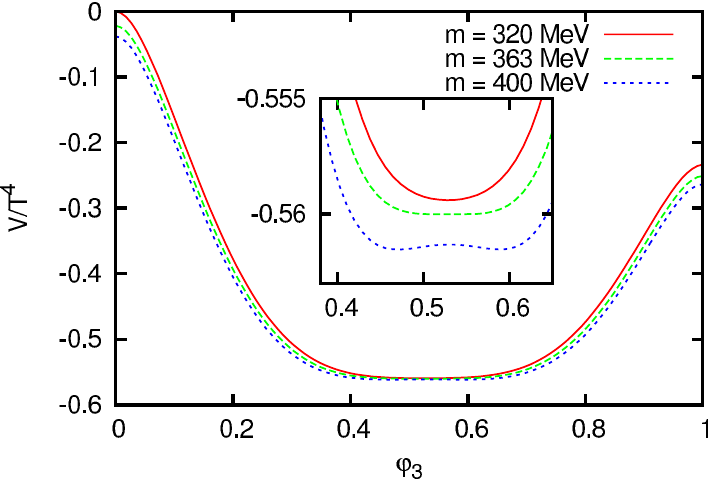}
\caption{The potential at $T_c$ for quark masses below, at and above $m_c=363$ MeV for
$N_f=1$ and $\mu=0$. The potentials have been shifted arbitrarily for better
visibility.\label{fig:potDiffMTc}}
\end{figure}
For smaller masses, $m<m_c$, we have a crossover. The related critical
temperature in this region is not unique and we extract it with the inflection point
$max(\partial_T L[\bar{A}_0])$.  The three different regions, first order
regime, critical point, crossover regime, are visualized in
Fig.~\ref{fig:potDiffMTc}, where we show the potential at the critical
temperature for three quark masses, which are below, at and above the
critical quark mass. For $m<m_c$ we find only one minimum away from
the confining value $\varphi_3=\frac{2}{3}$. At $m=m_c$ the potential
is flat. For $m>m_c$ we have two degenerate minima.
\begin{figure}[t]
\includegraphics[width=.45\textwidth]{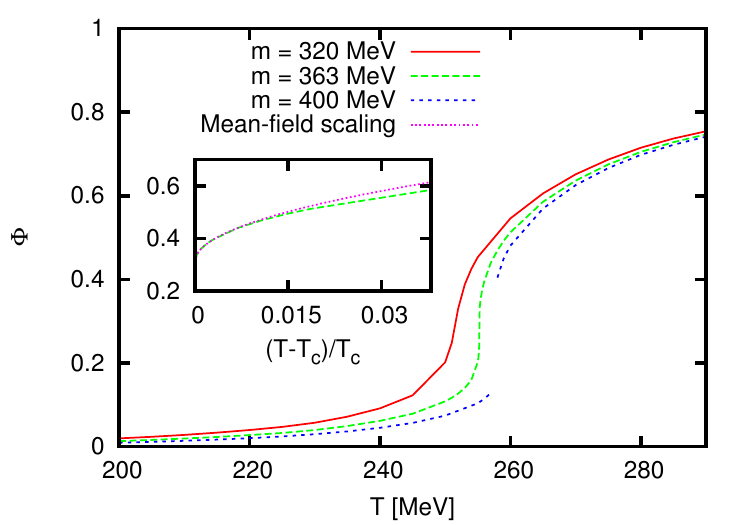}
\caption{The Polyakov loop $L[A_0]$ for the same quark masses as in 
Fig.~\ref{fig:potDiffMTc} for $N_f=1$ and $\mu=0$. \label{fig:plDiffM}}
\end{figure}
The structure of the potential reflects itself in the behaviour of the
order parameter, the Polyakov loop, which is shown in
\Fig{fig:plDiffM} as a function of $T$ for the same quark masses as
used for the potential in \Fig{fig:potDiffMTc}. One can clearly
distinguish the crossover for $m<m_c$ from the very sharp first order
transition at $m>m_c$. At $m=m_c$ we have a second order phase transition. The
critical exponent $\beta$ is derived from the order parameter with $L[\bar A_0]
\sim |T - T_c|^\beta$ in the vicinity of $T_c$. In our approximation we extract
$\beta=1/2$, the mean-field critical exponent. This is to be expected as the
current approximation to the Polyakov loop potential neglects the backreaction
relevant for criticality, see \cite{Marhauser:2008fz,Fister:2013bh}.

\begin{table}
\begin{tabular}{|c|c|c|c|}
\hline
$N_f$ & $m_c$ [MeV] & $m_c/T_c$ \\
\hline
1 & 363 & 1.422  \\
\hline
2 & 461 & 1.827  \\
\hline
3 & 509 & 2.038 \\
\hline
\end{tabular}
\caption{Critical current quark mass $m_c(\zeta=80\,\text{GeV})$.
\label{tab:McMu0}}
\end{table}

In Tab.~\ref{tab:McMu0} we list the resulting critical quark masses
for $N_f\in\{1,2,3\}$. Note that the current quark mass depends on the
renormalization point and scheme. Therefore a direct comparison of our results for
$m_c$ to the lattice results is difficult and will be postponed. We came back
to this point in section \ref{sec:results}. 

\subsection{Real chemical potential}

For real chemical potential and non-vanishing background fields the
quark contribution to the Polyakov-loop potential $V(\bar A_0)$ is
complex for general $(\varphi_3,\varphi_8)$. However, a modified real
effective potential can be constructed that agrees with $V(\bar A_0)$
on the equations of motion, \cite{JMP11}.  This procedure is
consistent with taking $\varphi_8=0$, leading to a real potential.
For this choice the same procedure to find $m_c$ as in the $\mu=0$
case can be used.  The chemical potential affects the Polyakov loop
potential dominantly through the quark loop in Fig.~\ref{fig:DSE-A},
but also through the back-reaction of the quarks onto the gluon, see
Fig.~\ref{fig:qgDSEs}.  We show the resulting critical quark masses in
Sec.~\ref{sec:results}.

\subsection{Imaginary chemical potential}
\begin{figure}[t]
\includegraphics[width=.45\textwidth]{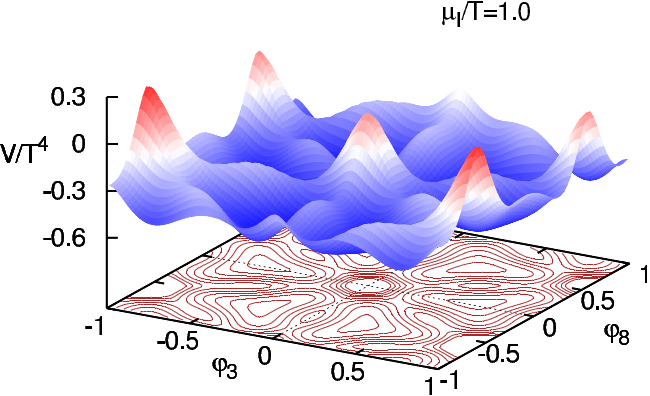}
\caption{Polyakov-loop potential at $\mu_I/T=1.0$, $T=250$ MeV, $m=320$ MeV. \label{fig:pot2dtheta}}
\end{figure}
At imaginary chemical potential $\mu_I=\mu/i=2\pi T\theta$ QCD features the
Roberge-Weiss symmetry \cite{Roberge:1986mm}, which states that QCD is periodic
in $\theta$ with periodicity $1/N_c$ defined by the gauge group.  Above the
Roberge-Weiss critical point, QCD exhibits first order phase transitions with
the Polyakov loop proportional to different center elements across the
transitions. In the functional approach, the physics of these transitions has
been studied in Ref.~\cite{Braun:2009gm}.  We built upon this framework here
and repeat the basic ideas in the following.

In the presence of an imaginary chemical potential, the quark Matsubara modes are
shifted by $\mu_I$ as well as $A_0$, the full gauge field. 
\begin{equation}
\omega_n \rightarrow \omega_n +g\, A_0 + 2 \pi T \theta\,. \label{eq:matsuShift1}
\end{equation}
For an Abelian gauge field $A_0$, the imaginary chemical potential
$\mu_I= 2\pi T\theta$ could be reabsorbed in the gauge field by a
shift leading to $g\,\tilde A_0 = g\, A_0 +2 \pi T \theta$. This is a
constant shift of the gauge field and does not change the gauge
action. For an $SU(N)$-gauge field this is not possible as it has no
$U(1)$-component.  However, we can use
\begin{equation}
\tilde\varphi_3 = \varphi_3 + 3\theta, \;\;\; \tilde\varphi_8 = \varphi_8 + \sqrt{3}\theta,
\end{equation}
and $\tilde{A}_0$ is the field with $\varphi_i$ replaced by $\tilde{\varphi}_i$.
With this shift, we can write Eq.~(\ref{eq:matsuShift1}) as
\begin{equation}
\omega_n\delta_{ij}+ g A_{0,ij} + 2 \pi T\theta\delta_{ij}\rightarrow
\omega_n\delta_{ij}+g \tilde{A}_{0,ij} + 3 (2 \pi T)\theta \delta_{i1}. \label{eq:matsuShift2}
\end{equation}
In this form the Roberge-Weiss symmetry is manifest: a shift in the imaginary
potential by $\theta \rightarrow \theta + \theta_z$ with $\theta_z=k/3$,
$k\in\mathbb{Z}$ can be absorbed by a shift in the Matsubara sum and a center
transformation in the gauge field.  As a center transformation is reflected by
a rotation of the Polyakov loop in the complex plane, the Polyakov loop must be
defined complex at imaginary chemical potential.

The minima of the Polyakov loop potential are therefore not at $\varphi_8=0$,
as the latter implies a vanishing $\arg(L[\bar{A}])$.  The potential for a
non-vanishing $\theta$ as a function of $\varphi_3$, $\varphi_8$ is shown in
Fig.~\ref{fig:pot2dtheta}. The minimum is shifted away from $\varphi_8=0$,
implying a complex Polyakov loop.  In contrast to the situation at real
chemical potential, the potential stays real at imaginary chemical potential
for all $\varphi_8$. 

\begin{figure}[t]
\includegraphics[width=.45\textwidth]{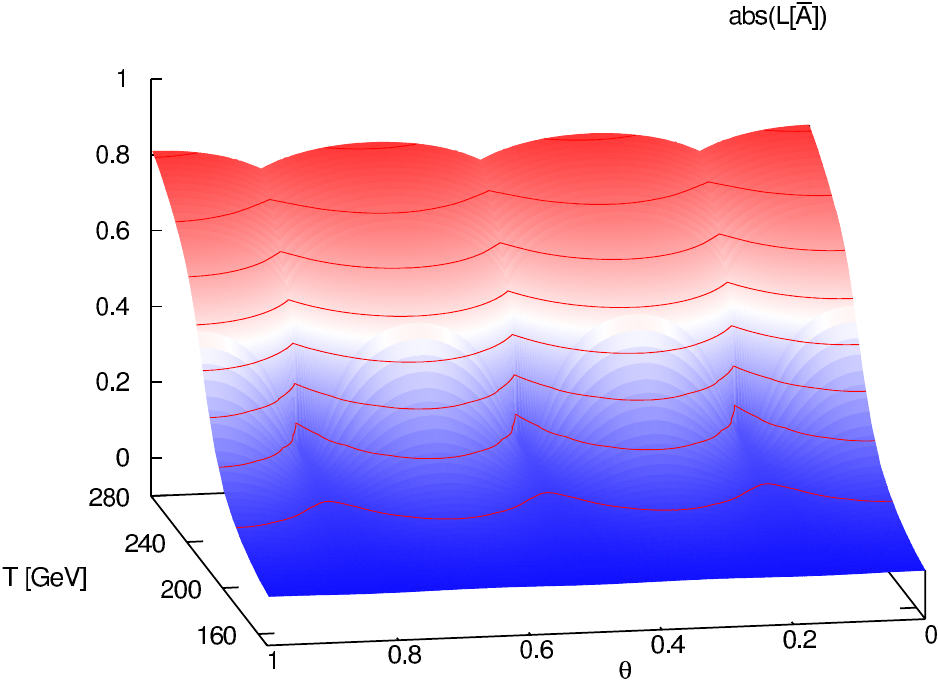}
\caption{Absolute value of the Polyakov loop at $m=200$ MeV for $N_f=2$. \label{fig:absL}}
\end{figure}

\begin{figure}[t]
\includegraphics[width=.45\textwidth]{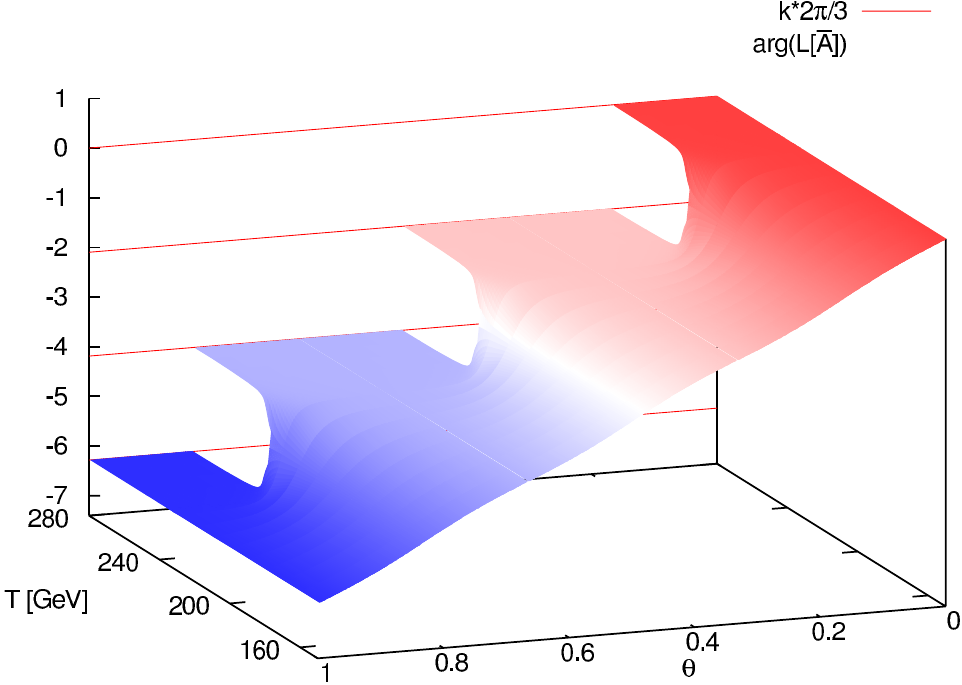}
\caption{Angle of the Polyakov loop at $m=200$ MeV for $N_f=2$. \label{fig:argL}}
\end{figure}

To demonstrate the effect of the imaginary chemical potential on the Polyakov
loop, we show the absolute value and the argument of the Polyakov loop,
respectively in Figs.~\ref{fig:absL},\ref{fig:argL} for $m=200$ MeV and
$N_f=2$.  The Roberge-Weiss symmetry is apparent. Above a critical temperature
of about $240$ MeV, $\arg(L[\bar{A}])$ shows a jump at
$\mu_I/T=\frac{\pi}{3}(2k+1)$, $k\in\mathbb{Z}$.  This is the Roberge-Weiss
phase transition, where the phase of the Polyakov loop expectation value jumps.
At asymptotically large temperatures it jumps from one center element to the
next.  At the value of the quark mass used for Fig.~\ref{fig:absL}, the
Polyakov loop shows a crossover at all values of $\theta$ up to the critical
value $\theta_c=1/6$. For a larger value of the quark mass, a critical
end-point appears for some $\theta$ at which the transition turns first order.
At the critical quark masses given in Tab.~\ref{tab:McMu0} this end-point is at
$\theta=0$.
\begin{figure}[t]
\includegraphics[width=.45\textwidth]{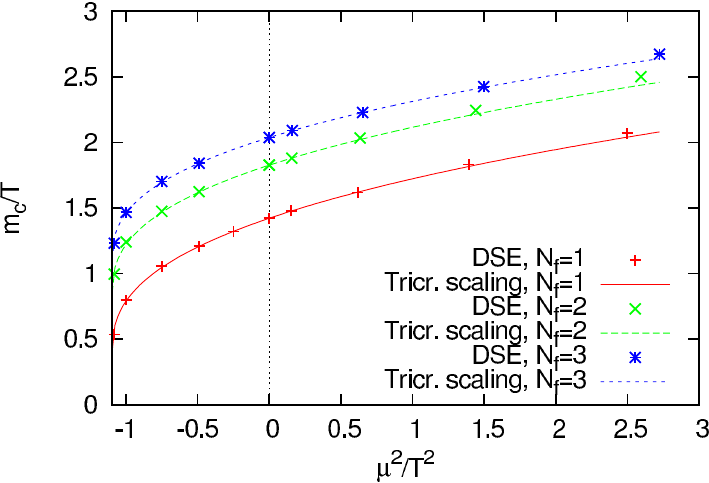}
\caption{The critical quark mass as a function of $(\mu/T)^2$. \label{fig:mcVsmu2Nf1}}
\includegraphics[width=.45\textwidth]{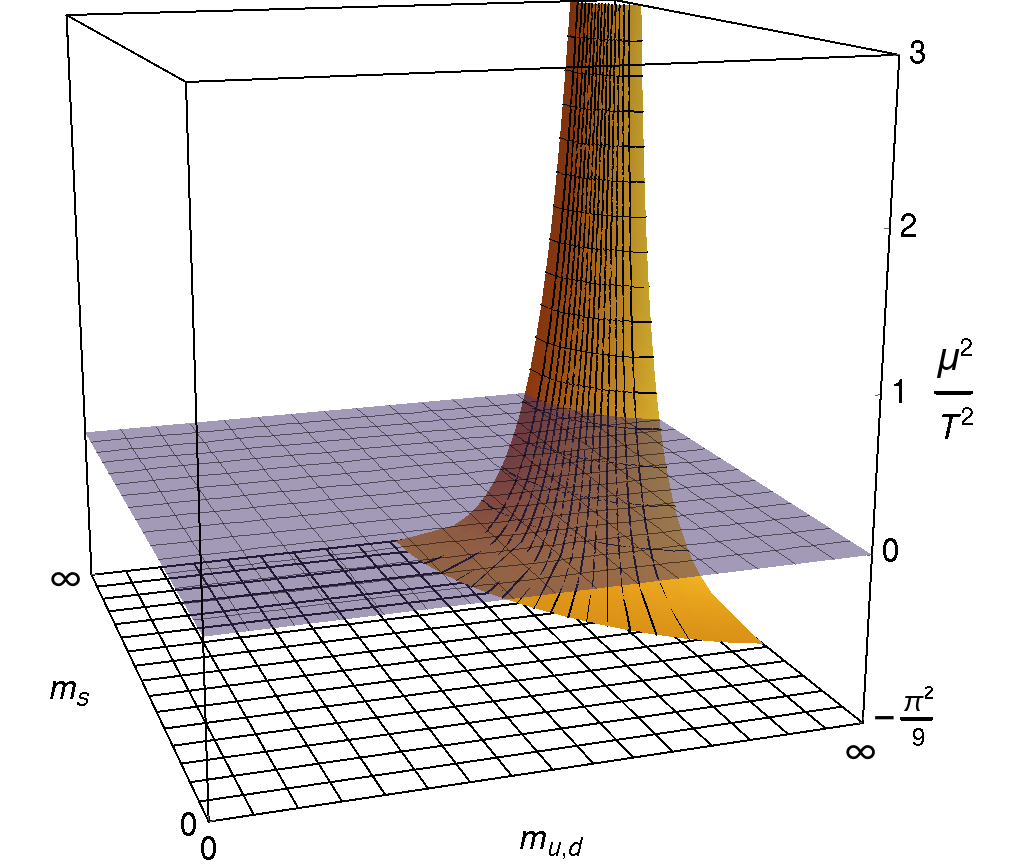}
\caption{The three-dimensional Columbia plot. The quark masses are scaled according to
$f:[0,\infty]\rightarrow[0,1],f(m)=1-e^{-m/T}$. \label{fig:columbia3d}}
\end{figure}

\section{Results for all chemical potentials \label{sec:results}}

The main result of this work is given in Fig.~\ref{fig:mcVsmu2Nf1},
where we show $m_c/T$ as a function of $\mu^2/T^2$ for $N_f\in\{1,2,3\}$
starting at the critical surface where $\mu/T=i\pi/3$.
We compare our results to tricritical scaling, given by
\begin{equation}
\frac{m_c}{T} = \frac{m_{tric}}{T} + K\left[\left(\frac{\pi}{3}\right)^2+
\left(\frac{\mu}{T}\right)^2\right]^{2/5}, \label{eq:tricScaling}
\end{equation}
where $m_{tric}$ is the quark mass on the tricritical surface $\mu_I/T=\pi/3$
and $K$ is a parameter that we determine by fitting.  In \cite{Fromm:2011qi} it
has been found that lattice results for $m_c/T$ are well described by 
tricritical scaling, Eq.~(\ref{eq:tricScaling}), up to large chemical
potentials.  As Fig.~\ref{fig:mcVsmu2Nf1} shows, we find the same agreement.
Only at very large $\mu^2$ we find slight deviations from the scaling behaviour
with our critical quark masses slightly above the scaling curve. This
has also been observed on the lattice \cite{Fromm:2011qi}. Thus we find
nice qualitative agreement with the lattice results at all chemical potentials.

\begin{table}
\begin{tabular}{|c|c|c|c|}
\hline
$N_f$ & $K$ & $m_{tric}/T$ \\
\hline
1 & 0.98 & 0.406  \\
\hline
2 & 0.94 & 0.852  \\
\hline
3 & 0.90 & 1.109 \\
\hline
\end{tabular}
\caption{Fit results for Eq.~(\ref{eq:tricScaling}).
\label{tab:KandMtric}}
\end{table}

The results for the parameters $K$ and $m_{tric}$ for the tricritical scaling
can be found in Tab.~\ref{tab:KandMtric}.  When we compare to those values
given in \cite{Fromm:2011qi} we find that our critical quark masses are
consistently smaller. This is certainly due to the different definition of 
the quark mass, which depends strongly on the renormalisation point and 
scheme and potentially also on the lattice volume. In order to make
quantitative statements one would have to compare appropriate quantities
as for example the corresponding quark propagators, see e.g. \cite{Fischer:2005nf} 
for such a comparison. Since these are not yet available from the lattice, 
we postpone this to future work.

In order to make the connection to the Columbia plot,
Fig.~\ref{fig:ColumbiaPlot}, we show in Fig.~\ref{fig:columbia3d} a
three-dimensional version of the Columbia plot, with $(\mu/T)^2$ as a third
axis. The first order area starts at the critical surface with
$(\mu/T)^2=-(\pi/3)^2$ and shrinks for growing $\mu^2$.  For
Fig.~\ref{fig:columbia3d} we also obtained the critical quark masses for
$N_f=2+1$ case with $m_l \neq m_s$ at $(\mu/T)^2\in\{-1.04^2,0,1.4^2\}$ and
continue these values by a fit for $m_{s,c}(m_{l,c})$ and tricritical scaling
for the $\mu$-dependence.

\section{Conclusions}\label{conclusions}

We have used an established truncation for the quark and gluon DSEs in order to
access the Polyakov-loop potential of full QCD at real and imaginary chemical
potential.  For large quark masses, we find an area where the phase transition
is of first order, bounded by a line of second order phase transitions. This is
the expected scenario from the Columbia plot.  As a function of $\mu^2$ the
critical quark masses are in qualitative agreement with lattice results. The
level of quantitative agreement is hard to access, since the definition of the
quark mass is different.

Given the quality of our description of the physics of confinement, it is a
logical extension to also investigate the physics of the chiral sector, i.e.
the lower left corner of the Columbia plot, Fig.~\ref{fig:ColumbiaPlot}.
There, the important physics is not the Polyakov-loop potential but the
Goldstone bosons of chiral symmetry. Since those are currently not included in
our truncation, we can not expect a valid description of this area so far.
However, it is possible to include the back-reaction of the Goldstone bosons
onto the quarks in the DSE language, see e.g. \cite{Fischer:2011pk}.  An
extension of this work will therefore be able to describe the critical physics
of confinement and chiral symmetry breaking at all quark masses and chemical
potentials.

\section*{Acknowledgements}

We would like to thank Leonard Fister for dicusssions and work on
related projects and Bernd-Jochen Schaefer for a critical reading
of the manuscript. This work is supported by the Helmholtz Alliance
HA216/EMMI, by ERC-AdG-290623, by the Helmholtz International Center
for FAIR within the LOEWE program of the State of Hesse and by the
BMBF grant 05P12VHCTG.

\bibliography{hqpapers.bib}

\end{document}